Phonon study of Jahn-Teller distortion and phase stability in NaMnO$_2$ for sodium-ion batteries

Haeyoon Jung[1], Jiyeon Kim[1,2*], Sooran Kim[1*]

[1]Department of Physics Education, Kyungpook National University, Daegu 41566, South Korea

[2]The Center for High Energy Physics, Kyungpook National University, Daegu 41566, South Korea

*Corresponding authors: mygromit@gmail.com, sooran@knu.ac.kr

**Abstract**

Cathode materials undergo various phase transitions during the charge/discharge process, and the structural transitions significantly affect the battery performance. Although phonon properties can provide a direct clue for structural stability and transitions, it has been less explored in sodium cathode materials. Here, using the first-principles calculations, we investigate phonon and electronic properties of various layered $NaMnO_2$ materials, especially focusing on the dependency of the Jahn-Teller distortion of $Mn^{3+}$. The phonon dispersion curves show that the O'3 and P'2 structures with the Jahn-Teller distortion are dynamically stable in contrast to undistorted O3 and P2 structures. The structural instability of O3 and P2 structures is directly observed from the imaginary phonon frequencies, as so-called phonon soft modes, whose corresponding displacements are from O atoms distorting along the local Mn-O bond direction in the $MnO_6$ octahedra. This is consistent with the experimental stability and a structural transition with the Jahn-Teller distortion at the high Na concentration. Furthermore, the orbital-decomposed density of states presents the orbital redistribution by the Jahn-Teller distortion such as $e_g$-band splitting, and the stability of O'3 and P'2 is not sensitive to the electron-electron correlation. Our results demonstrate the importance of phonon analysis to further understand the structural stability and phase transitions in cathode materials.

**I. Introduction**

Rechargeable lithium-ion batteries (LIBs) have been intensively and extensively studied as energy storage for electronic devices due to their high energy density, high capacity, and long cycle life performance.[1,2] Despite high demands for LIBs, limited supply and uneven distribution of lithium cause high cost,[3] which promotes a search for new materials to replace LIBs. Sodium is considered a good alternative for lithium because of its similar chemical properties in the same group and its natural abundance. Sodium-ion batteries (SIBs) thus have drawn a lot of interest as a promising substitute for LIBs, especially for large-scale energy storage applications. Layered transition metal oxides, such as Na$Me$O$_2$ ($Me$= 3$d$ transition metal elements) are one of the most investigated cathode materials for SIBs because of their simple structure and diversity of materials upon synthesis methods[4–9]. A broad choice of transition metal ions from Ti to Cu in Na$Me$O$_2$ provides a flexible and various design for advanced electrodes compared to the lithium case that has rather limited combinations of Co, Mn, and Ni.[10–12]

Among layered Na$Me$O$_2$, Mn-based oxides are attractive candidates for large-scale batteries due to their abundance and safety. Na$_x$MnO$_2$ exhibits various phases depending on synthesis conditions.[13–16] Na-layered oxides have been classified by Delmas *et al.*[17], and the representative structures are O3-type and P2-type structures as shown in Figure 1. O and P denote the octahedral and prismatic sites where Na is located, respectively, while the numbers (3 and 2) indicate the number of different oxygen packing (AB, BC, CA, and AB, BA) in MnO$_6$ layers. O3- and P2-type Na$_x$MnO$_2$ both exhibit the high capacity of 197 mAh/g[18] and 216 mAh/g,[19] respectively. Furthermore, distorted structures of O3- and P2-types (denoted as O'3 and P'2, respectively) are thermodynamically stable at high Na content (x > 0.67) with the Jahn-Teller (J-T) distortion of $Mn^{3+}$ ion.[18,19] The P2-P'2 phase transition triggered by the J-T active $Mn^{3+}$ ion was observed at the low voltage.[20–22]

The J-T distortion and oxidation state of Mn in layered $NaMnO_2$ are important factors in battery performance. The octahedral environment around Mn ion splits $d$ orbitals into $t_{2g}$ and $e_g$ orbitals, and $3d^4$ electrons of $Mn^{3+}$ ion gives to $t_{2g}^3$ and $e_g^1$ configurations in a high spin state. $Mn^{3+}$ in the octahedra environment typically accompanies the strong J-T effect, which lowers the energy by splitting the triply and doubly degenerate $t_{2g}$ and $e_g$ states, with the Mn-O bond length distortion as in Figure 1(e).[23] Various experimental and theoretical studies on $NaMnO_2$ have been reported focusing on the J-T effect and Na/Mn contents.[24–35] Experimentally, for example, P2-type $Na_{0.6}MnO_2$ exhibits the J-T distortion by inserting $Na^+$ ion and associated-structural change during the $Na^+$ extraction/insertion process that results in poor reversibility.[27] The J-T distortion can be suppressed by Mg doping in $Na_{0.67}MnO_2$ and the system shows smooth charge/discharge profile and improved capacity retention.[24] Furthermore, cooperative J-T distortion and Na ordering emanate a superstructure in $Na_{5/8}MnO_2$, which couples to long-range magnetic order at low temperature.[25] Theoretically, the electronic structures and energy gain from the J-T distortion were reported using first-principles calculations.[28,33–35] From the density of states of O'3-type $NaMnO_2$, the $t_{2g}$ states of Mn-$d$ orbitals contribute to the lower and middle regions of the valence band, and the top of the valence band is occupied by $d_{3z^2-r^2}$ states. The energy difference between $t_{2g}$ and $e_g$ ($d_{3z^2-r^2}$) is relatively small, and Mn ions favor the high spin configuration.[35] For orthorhombic P2-type, $Mn^{4+}$ and $Mn^{3+}$ orderings were investigated from the formation energy calculations, and the activation energy for Na diffusion varies from 273 meV to 423 meV during the $Na^+$ intercalation/deintercalation process.[29] In addition, the Coulomb interaction in layered Li/Na cathode materials including O'3 and P'2-$NaMnO_2$ are examined using the constrained random phase approximation (cRPA) method, where the Coulomb interaction parameters for localized orbitals are mainly determined by the local structure rather than a global structure features such as different stacking types.[36]

Despite several computational studies on energetics and electronic structures of $NaMnO_2$ and its J-T distortion,[28–35] phonon properties of layered $NaMnO_2$ have been less focused and explored using the first-principles calculations. There have been a few phonon studies on Li-ion cathode materials.[37–39] However, phonon properties of Na-ion cathode materials, especially their relation to structural transitions, have been less under scrutiny. Since phonon study can analyze the dynamic stability of a material and directly demonstrate structural phase transitions, it would be worth investigating the phonon character of cathode materials that undergo structural transitions during the charge/discharge process.

Here, we present the phonon/lattice dynamics and the electronic properties of various layered $NaMnO_2$. Figure 1 illustrates the crystal structures of $NaMnO_2$ systems studied in this work. In the phonon band structures, we observe the contribution of the J-T distortion to the dynamical stability of O'3 and P'2-type structures and instability of O3 and P2-types. The phonon soft modes are related to the J-T like displacements, which is consistent with the experimental structure and phase transition at the high Na concentration. In addition, we explore the stability of O'3 and P'2 depending on the strength of on-site Coulomb interaction. Our study shows that not only energetics but also phonon analysis provides important insights to understand the structural stability of cathode materials.

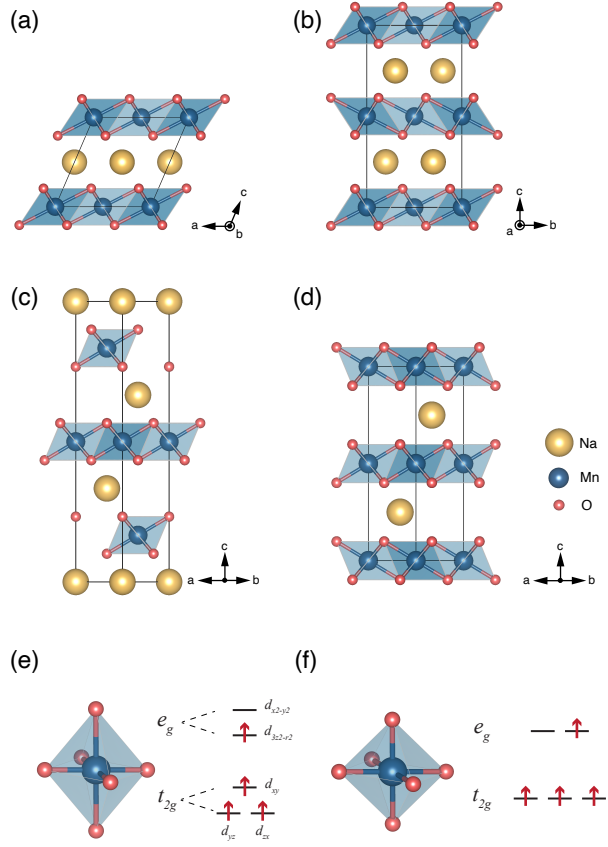

Figure 1. Various crystal structures of NaMnO$_2$. (a) O'3-type with monoclinic structure (space group (s.g.) *C2/m*). (b) P'2-type with orthorhombic structure (s.g. *Cmcm*). (c) O3-type with trigonal structure (s.g. $R\bar{3}m$). (d) P2-type with hexagonal structure (s.g. *P6$_3$/mmc*). (e,f) Local MnO$_6$ octahedra structures and electronic configurations (e) with and (f) without the J-T distortion.

## II. Computational Methods

To understand the lattice dynamics and electronic structures of NaMnO$_2$, we performed the spin-polarized density-functional-theory (DFT) calculations using the Vienna *ab-initio* simulation package (VASP).[40,41] All calculations were carried out using Perdew-Burke-Ernzerhof (PBE) of the generalized gradient approximation (GGA) as an exchange-correlation functional.[42] To access the localized Mn-3*d* orbital, we considered the effective on-site Coulomb interaction parameter, $U_{eff} = U-J = 3.9$ eV unless specified otherwise.[43] The various $U_{eff}$ values(0.0, 1.0, 2.0, 3.24, 3.62, 3.9, 4.0, 5.0 eV) are also used for a detailed investigation of its effect on structural stability. $U_{eff}$=3.24 eV and 3.62 eV were reported for O'3-NaMnO$_2$ and P'2-NaMnO$_2$, respectively, from the cRPA calculations.[36] A plane-wave energy cutoff of 520 eV was employed, and the *k*-point density was 5000/atom. The monoclinic O'3, orthorhombic P'2, two hexagonal O3 and P2-type structures are adopted whose space groups are *C*2/*m*, *Cmcm*, $R\bar{3}m$, and *P*6$_3$/*mmc*, respectively.[22,44–46] The corresponding *k*-point grids for the density are 6×13×6, 12×6×3, 13×13×2, and 13×13×3 *k*-grid including the Γ point for O'3, P'2, O3, and P2-NaMnO$_2$, respectively. We fully relaxed the atomic coordinates and the lattice parameters until the Hellmann-Feynman force was reduced to less than 0.01 eV/Å.

For phonon dispersion calculations, we employed PHONOPY package where the force constants and dynamic matrix are obtained from the supercell approach with finite displacements.[47] The supercell size of O'3, P'2, O3, and P2 structures was 2x2x2 for the phonon calculations.

**III. Results and Discussion**

Figure 2 shows the phonon dispersion curves along with the high symmetry directions of the Brillouin zone and phonon density of states (DOS) of O'3, P'2, O3, and P2 structures. Both O'3 and P'2 phases in Fig. 2(a) and (b) are dynamically stable since imaginary frequencies, so-called phonon soft modes, are not observed in the phonon bands. The stable phonon dispersions of O'3 and P'2 are consistent with the experimental observation that both O'3 and P'2 structures can be synthesized under high Na content.[18,19] On the other hand, the O'3 structure is energetically more favorable than the P'2 structure by ~59 meV/f.u. in the total energy calculations. Namely, P'2 is unstable compared to O'3 in the point of energy while P'2 is dynamically stable according to phonon calculations. A combination of phonon and total energy calculations provides the proper description for the P'2 structure that it exists as a metastable structure.

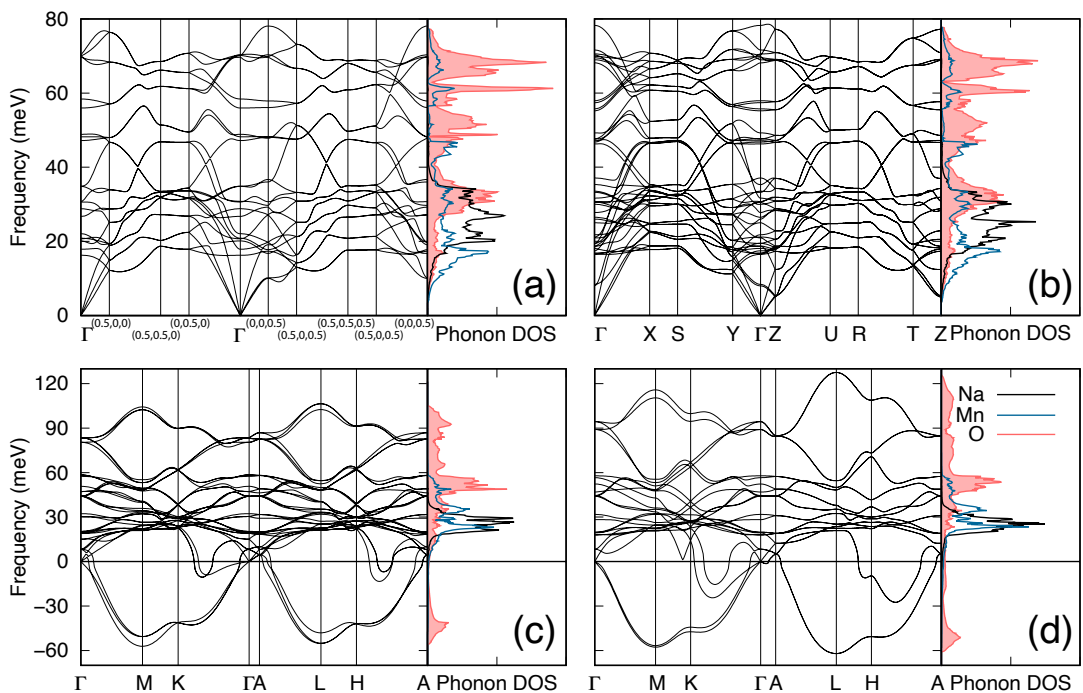

Figure 2. Phonon dispersion curves and phonon density of states of (a) O'3 (b) P'2 (c) O3, and (d) P2-type NaMnO$_2$. The imaginary phonon frequencies indicate the structural instability.

The similar phonon DOSs for O'3 and P'2 indicate that layer stacking does not significantly affect the phonon distribution. The displacements of Mn atoms mainly occur at low frequencies, while those of O atoms dominate at a high frequency about above 50 meV region due to their atomic mass. Na phonon bands contribute generally at the medium frequency region. This distribution of lattice vibrations of NaMnO$_2$ is consistent with the phonon DOS of LiMnO$_2$.[37]

In contrast to O'3 and P'2 cases, the phonon structure of O3 and P2 in Fig. 2(c) and (d) exhibit imaginary phonon frequencies at the *M*, *K*, *L*, and *H* points. These phonon soft modes reflect the structural instability, showing the essential role of the J-T distortion in structural stability. The dynamically stable O'3 and unstable O3 from the phonon calculations are consistent with the experimentally stable O'3-NaMnO$_2$. The stable P'2 and unstable P2 in the phonon bands properly reproduce the structural P2-P'2 phase transition at the low voltage during the discharge process. As in P'2 and O'3, the phonon DOS of O3 and P2 exhibit the oxygen contribution at the high frequency region and Mn and Na at the low and medium frequency region. It is worth noting that the phonon soft modes in Fig. 2(c) and (d) are mostly related to the lattice vibrations of O atoms, which can result in the distortion of the MnO$_6$ octahedra. We will discuss below that the displacement of the phonon soft modes can be responsible for the structural transition with the J-T type distortion.

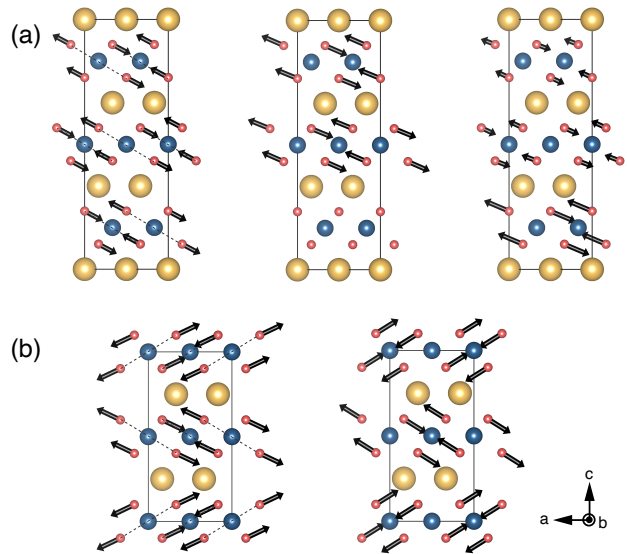

Figure 3. Atomic displacements corresponding to the phonon soft modes at the *M* point of (a) O3-type, and (b) P2-type. They are presented in increasing frequency order from left to right. Blue, red, and yellow balls indicate Mn, O, and Na ions, respectively.

Figure 3 illustrates the representative atomic displacements of the phonon soft modes of O3 and P2 type NaMnO$_2$ at the *M* point where the lowest phonon frequency is observed in the O3 type structure. O3-NaMnO$_2$ exhibits three phonon soft modes: one lowest and two degenerated second-lowest modes (Fig. 2(c)) while the P2 structure has two phonon soft modes (Fig. 2(d)) at the *M* point. All these modes are associated with the lattice vibrations of O atoms as shown in Fig. 3 and phonon DOS of Fig. 2(c,d).

In the intralayer, the displacements of the oxygens are mainly elongation or compression along the Mn-O bond axis (dashed lines in Figure 3). Note that this kind of adjustment of the Mn-O bond length in the MnO$_6$ octahedra is the main structural displacement of the J-T distortion. The displacement is also consistent with the distortions activating collective cluster deformations $Q_4$ and $Q_5$, which can describe the J-T distortion.[28] All phonon modes exhibit similar Mn-O bond length distortions in the intralayer for both O3 and P2 structures but only differ by the ordering of the intralayer motion among MnO$_6$ layers. In addition, the phonon soft modes at other *q*-points, such as the *K*, *L*, and *H*, also have the similar intralayer atomic movements from the oxygen atoms resulting in the lengthened or contracted Mn-O bond. The variances among the displacements at *q*-points come from the sequence of intralayer displacement along the *c* direction as at the *M* point. These results show that the lattice displacement from the phonon soft modes appropriately describes the phase transition with J-T distortion from P2 to P'2 and structural transform from O3 to O'3.

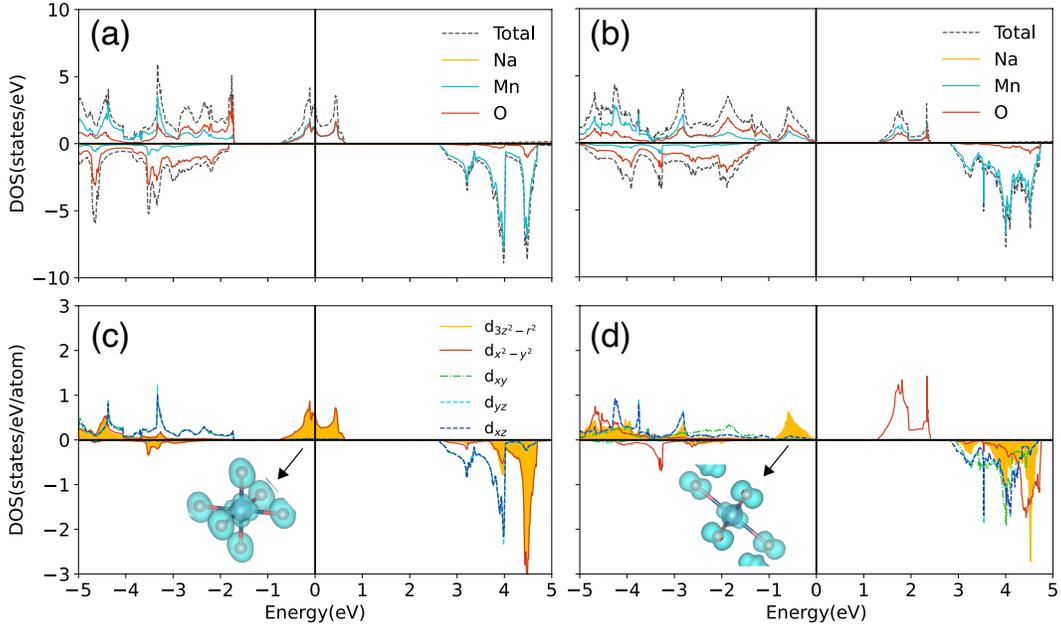

Figure 4. Total and atomic density of states in (a) O3 and (b) O'3-type structures. Projected density of states (PDOS) of Mn $3d$ orbitals for (c) O3 and (d) O'3-types. (Inset) Charge density isosurface of $MnO_6$ octahedra.

To investigate the J-T effect on the electronic properties, we have compared the electronic DOS with and without the distortion. Figure 4 shows the total and partial DOS of O3 and O'3 structures. The valence bands near the Fermi level in both O3- and O'3-$NaMnO_2$ are mainly contributed by O-$2p$ and Mn-$3d$ states as shown in Figures 4(a) and (b). These total and atomic DOSs of the distorted O'3 structure are consistent with previous reports.[33–35] Spin-polarized calculations show high spin states of $Mn^{3+}$, separated into up and down spin states according to spin-exchange energy. The calculated magnetic moment of Mn in O'3 and O3 are 3.801 $\mu_B$ and 4.02 $\mu_B$, respectively. The J-T distortion significantly alters the DOS by opening the band gap of 1.28 eV as shown in Fig. 4(b) and (d). By opening the band gap via J-T distortion, the energy gain of 0.355 eV/f.u., which is the energy difference between O'3 and O3, is obtained.

We further explored the $d$ orbital decomposed DOS of O3 and O'3 as in Fig. 4(c) and (d), respectively, where the elongated bond of Mn-O is aligned to the local $z$-direction. In the case of O3, one electron is evenly distributed in the doubly degenerated $e_g$ states with metallic character. The three $t_{2g}$ orbitals are degenerated as in occupied band below -1.7 eV and unoccupied band above 2.6 eV. The energy gap between $t_{2g}$ and $e_g$, so-called 10$Dq$, is 0.96 eV. On the other hand, the O'3 structure with the J-T distortion exhibits the splitting of $e_g$ states into $d_{3z^2-r^2}$ and $d_{x^2-y^2}$ by the asymmetric crystal field, which in turn opens a gap by fully filling the lower bands of $d_{3z^2-r^2}$ as in Fig. 4 (d). The elongated bond of Mn-O along the local $z$-direction lifts the $d_{x^2-y^2}$ states. We also found the splitting of occupied $t_{2g}$ bands that lowers degenerated $d_{yz}$, $d_{xz}$, and increases $d_{xy}$ in energy compared to the O3 structure. The gap between $t_{2g}$ and $e_g$, 10$Dq$, is almost closed by increasing $d_{xy}$ of $t_{2g}$ bands and lowering $d_{3z^2-r^2}$ of $e_g$ bands through the J-T effect. Furthermore, the charge density isosurfaces in the inset of Fig. 4 clearly show the orbital redistribution by the J-T distortion. The directional bonding between Mn $d_{3z^2-r^2}$ and O $p_z$ is observed in O'3-NaMnO$_2$.

In addition, the P'2 structure exhibits the splitting of $e_g$ level and band gap of 1.23 eV with J-T distortion, which results in lowering the energy by 0.350 eV/f.u. compared to the P2 structure. The similar DOS and energy gain of the P'2 structure in comparison to those of O'3-NaMnO$_2$ suggest that layer stacking sequence does not significantly adjust the detailed electronic structures or lattice stability rather than the local distortion effect.

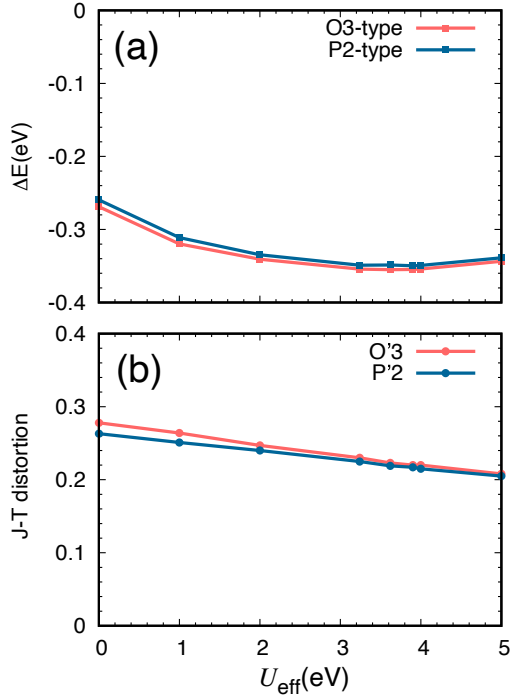

Figure 5. $U$ dependence of the energy gain and J-T distortion. (a) Total energy differences of O-type ($E$(O'3) − $E$(O3)) and P-type ($E$(P'2) − $E$(P2)) structure. (b) The magnitude of the J-T distortion in O'3 and P'2 structures.

To investigate the role of the electron correlation in lattice stability and the J-T distortion, we calculated the energy and magnitude of the J-T distortion depending on $U_{\text{eff}}$. Figure 5(a) shows the total energy difference between distorted (O'3 and P'2) and undistorted (O3 and P2) structures as a function of $U_{\text{eff}}$. The total energy of the O'3 (P'2) structure is lower than that of the O3 (P2) structure in all studied $U$ values as shown as the negative energy difference Δ$E$. Namely, the J-T distortion is preferable regardless of the strength of the electron correlation. The energy difference Δ$E$ between O'3 and O3 (or P'2 and P2) becomes larger up to $U_{\text{eff}}$ of 4 eV and slightly smaller after 4 eV, which is consistent with the case of NaNiO$_2$.[28] The Δ$E$ of O3(P2)-type with $U_{\text{eff}}$=0eV and 3.9 eV are -0.269(-0.260) eV and -0.355(-0.350) eV, respectively. Namely, the additional change of $\Delta E$ of O3(P2)-type by the inclusion of $U$ is -0.086(-0.090) eV at $U_{eff}$=3.9eV, which is much smaller than the energy gain of ~-0.269(-0.260) eV only with the J-T distortion. It suggests that the change of $\Delta E$ upon $U_{eff}$ is not significant, and the lattice stability is not considerably altered by the Coulomb correlation.

To examine the J-T distortion as a function of $U_{eff}$, the magnitude of the J-T distortion is evaluated as in (1),

$$\text{JT distortion} = \frac{6(l_{long} - l_{short})}{2l_{long} + 4l_{short}} \quad (1)$$

where $l_{long}$ and $l_{short}$ are the long bond and short bond of Mn-O in MnO$_6$ octahedra, respectively.[48] For example, with $U_{eff}$ = 3.9 eV, two long (four short) bond lengths of Mn-O in O'3 and P'2 is 2.436 (1.966) Å and 2.425 (1.965) Å, respectively, whose magnitude of the J-T distortion is 0.221 for O'3 and 0.217 for P'2. The J-T distortion magnitudes of O'3 and P'2 structures slightly decrease as the $U_{eff}$ increases as shown in Figure 5(b). We additionally calculated the energy gain, $\Delta E$ using the same $U_{eff}$ value ($U_{eff}$=0eV) with the relaxed structures from different $U_{eff}$ values to separately investigate how much this slight change of the J-T distortion affects the energy gain. The obtained $\Delta E$ is about -0.265 eV regardless of different relaxed structures, which is close to the $\Delta E$ value at $U_{eff}$=0. Namely, the difference in the magnitude of J-T distortion among various $U_{eff}$ is negligible, which indicates that the electron correlation does not affect the J-T distortion significantly.

**IV. Conclusion**

In conclusion, we investigated phonon and electronic properties of various layered NaMnO$_2$ with and without the J-T distortion using first-principles calculations. The phonon calculations provide the dynamic stability of the distorted structures, O'3 and P'2. Despite the higher energy of P'2 than that of O'3, the stable phonon bands of P'2 suggest that P'2 phase can exist as a metastable structure. The observation of dynamic stability and metastable structure of P'2-type implies the importance of phonon calculations for a comprehensive understanding of the stability of crystal structures. On the other hand, undistorted structures O3 and P2 exhibit the phonon soft modes indicating structural instability, which is in agreement with previous experimental results. The soft modes in O3 and P2 are related to the movement of O atoms along the local axis of the MnO$_6$ octahedra, which gives the possibility of the transition to a structure with the J-T distortion like O'3 and P'2. Thus, our phonon calculations appropriately describe the experimental stability of the O'3 and P2-P'2 phase transition at the low voltage. We also found that the strength of the Coulomb correlation does not significantly modify the lattice stability or J-T distortion. We believe that this work demonstrates the utility of the phonon approach providing insight to understand the stability/instability of cathode materials and various structural transitions during the charge/discharge process.

## Acknowledgements

We thank Bongjae Kim, Minjae Kim, and Jaewook Kim for helpful discussions. This work was supported by the Korea Electric Power Corporation (Grant R20XO02-12) and KISTI Supercomputing Center (Project No. KSC-2020-CRE-0255).

## Author Declarations

### Conflict of Interest

The authors have no conflicts to disclose.

### Data availability

The data that support the findings of this study are available from the corresponding author upon reasonable request.